\documentclass{jpp}

\usepackage{graphicx}
\usepackage{natbib}
\usepackage{bm}
\usepackage{epsfig}
\usepackage{amsmath}
\usepackage{amsfonts}
\usepackage{amssymb}
\usepackage{epstopdf}
\usepackage{lineno}
\usepackage[dvipsnames]{xcolor}
\usepackage[english]{babel}

\newcommand\etal{\mbox{\textit{et al.}}}

\newsavebox{\astrutbox}
\sbox{\astrutbox}{\rule[-5pt]{0pt}{20pt}}

\newcommand{\bB}{\mathbf{B}}
\newcommand{\bv}{\mathbf{v}}

\newcommand{\bE}{\mathbf{E}}
\newcommand{\bj}{\mathbf{j}}
\newcommand{\br}{\mathbf{r}}
\newcommand{\bzpm}{\mathbf{z^\pm}}

\newcommand{\bzp}{\mathbf{z^+}}
\newcommand{\bzm}{\mathbf{z^-}}

\usepackage{changes}
\definechangesauthor[name={Luca}, color=red]{LSV}
\definechangesauthor[name={Denise}, color=blue]{DP}
\definechangesauthor[name={Oreste}, color=Green]{OP}
\definechangesauthor[name={Sergio}, color=Maroon]{SS}
\definechangesauthor[name={Franco}, color=Gray]{FV}

\title[ Local energy transfer rate and kinetic processes]{Local energy transfer rate and kinetic processes: the fate of turbulent energy in two-dimensional Hybrid Vlasov-Maxwell numerical simulations}

\author[L. Sorriso-Valvo et al.]{Luca Sorriso-Valvo$^1$\thanks{Email address for correspondence: 
lucasorriso@gmail.com}, Denise Perrone$^{2,3}$, Oreste Pezzi$^4$, Francesco Valentini$^4$, Sergio Servidio$^4$, Ioannis Zouganelis$^2$ \& Pierluigi Veltri$^4$}
\affiliation{
$^1$Nanotec/CNR, Sede di Cosenza, Ponte P. Bucci, Cubo 31C, 87036 Rende, Italy.\\ 
$^2$European Space Agency, ESAC, Madrid, Spain. \\ 
$^3$Department of Physics, Imperial College London, London SW7 2AZ, United Kingdom.\\
$^4$Dipartimento di Fisica, Universit\`a della Calabria, Ponte P. Bucci, Cubo 31C, 87036 Rende, Italy.}

\pubyear{}
\volume{}
\pagerange{}
\date{?; revised ?; accepted ?. - To be entered by editorial office}

\begin{document}
\maketitle
 
\begin{abstract}
The nature of the cross-scale connections between the inertial range turbulent energy cascade and the small-scale kinetic processes in collisionless plasmas is explored through the analysis of two-dimensional Hybrid Vlasov-Maxwell numerical simulation (HVM), with $\alpha$ particles, and through a proxy of the turbulent energy transfer rate, namely the Local Energy Transfer rate (LET). 
Correlations between pairs of variables, including those related to kinetic processes and to deviation from Maxwellian distributions, are first evidenced. Then, the general properties and the statistical scaling laws of the LET are described, confirming its reliability for the description of the turbulent cascade and revealing its textured topology.   
Finally, the connection between such proxy and the diagnostic variables is explored using conditional averaging, showing that several quantities are enhanced in the presence of large positive energy flux, and reduced near sites of negative flux. These observations can help determining which processes are involved in the dissipation of energy at small scales, as for example ion-cyclotron or mirror instabilities typically associated with perpendicular anisotropy of temperature.  
\end{abstract}

\begin{PACS}
Authors should not enter PACS codes directly on the manuscript, as these must be chosen during the online submission process and will then be added during the typesetting process (see http://www.aip.org/pacs/ for the full list of PACS codes)
\end{PACS}

%\keywords{Vlasov, turbulence, dissipation}

%\added[id=LSV]{}

\section{Introduction}
\label{sec:intro}

%\item  Heating and Turbulence in plasmas $\rightarrow$ Connection between inertial and kinetic scales
The dynamical behavior of plasmas depends markedly on frequencies and scales. At low frequencies and large scales, ions and electrons are locked together and behave like an electrically conducting fluid; this is the regime of magnetohydrodynamics (MHD). At higher frequencies and smaller scales, electrons and ions can move relative to each other, behaving like two separate and interpenetrating fluids: this is the two-fluid regime. However, if the dynamics is supported by non-thermal effects and anisotropies in the velocity space, the more complete kinetic theory shall be adopted.  A notable example is represented by the solar wind, a continuous, highly variable, weakly collisional plasma outflowing at high speed from the Sun. This complex medium is highly structured, with turbulent fluctuations on a wide range of scales, from fractions of a second to several hours, thus covering all the above dynamical ranges.

Spacecraft measurements have revealed that the solar-wind plasma is usually in a state of fully developed turbulence~\citep{marsch06,bruno13,alexandrova13}. As the turbulent fluctuations expand away from the Sun, these decay and transfer energy to smaller scales. At scales comparable with the ion gyro-radius or the ion inertial length, the MHD approximation fails and kinetic processes involving field-particle interactions take place, along with important cross-scale coupling~\citep{crossscale}. Furthermore, since the plasma collisionality is quite low, the ion and electron velocity distribution functions can exhibit significant non-Maxwellian signatures. 
Kinetic processes manifest through the excitation of plasma waves, temperature anisotropy, plasma heating, particle energization, entropy cascade, excitation of plasma waves and so on \citep{vaivads16}. Recently, a phenomenological description in terms of cascade in the velocity space, due to highly structured particle distributions, has been proposed~\citep{scheko,servidio17}. Observations based on the high-resolution ion velocity distribution measurements by the Magnetospheric Multiscale (MMS) mission in the Earth's turbulent magnetosheath~\citep{servidio17} have recently confirmed such description, opening a new pathways to the understanding of plasma turbulence and dissipation in collisionless plasmas. This cascade may efficiently transfer the energy towards smaller velocity scales, thus producing finer structures in the particle velocity distribution function. The presence of such structures may ultimately enable and enhance the effect of collisions in dissipating the energy~\citep{pezzi14,pezzi15a,pezzi15b,pezzi16a,pezzi16b,pezzi17SW}. Numerical simulations represent an indispensable tool for the understanding of plasma nonlinear dynamics. The use of kinetic numerical simulations is crucial, indeed, when kinetic processes occur in turbulent collisionless plasmas \citep{parashar09,camporeale11,servidio15,franci15,cerri16,cerri17,groselj17,franci18}. The Eulerian hybrid Vlasov-Maxwell  {\it HVM} code~\citep{valentini07,perrone11} can be used to study the dynamics of a turbulent plasma with typical solar wind conditions, in order to describe the cross-scale coupling between the MHD turbulence and the small-scale kinetic processes~\citep{valentini08,valentini09,valentini10,valentini11,valentini11apj,perrone11,servidio12,greco12,perrone13,servidio14,perrone14a,perrone14b,valentini14,servidio15, valentini16, pezzi17a, pezzi17b}.

In turbulence, at small scales, the energy level of the fluctuations becomes too low to be investigated with poorly-resolved Lagrangian algorithms with very low number of particles-per-cell (for a convergence study, see~\citet{HaggertyEA17}). Eulerian-based numerical algorithms might be more suitable for this scope. HVM simulations show that kinetic effects produce, close to reconnection regions, a distortion of the ion velocity distribution, through the generation of accelerated field-aligned beams, temperature anisotropy and trapped particle populations. Moreover, departures from the Maxwellian shape are more pronounced for heavy ions than for protons~\citep{perrone13,perrone14b,valentini16}, pointing out that the role of secondary ions must be considered for understanding energy partition in plasmas~\citep{marsch06,kasper08,araneda09,maneva15}. Although the physical mechanisms of solar wind turbulence have been matter of investigation for many decades, both using `in situ' measurements and numerical experiments, important problems concerning the interconnection between the inertial and kinetic range processes, as well as the interplay between physical space fluctuations and velocity space perturbations, are still poorly understood. 

%\item  Intermittency in SW $\rightarrow$ Structures: LIM, PVI
%\item  Scalogram Toti: need for more than just PVI $\rightarrow$ more complete estimate of energy including v, v.b...
Solar wind turbulence is characterized by intermittency~\citep{Burlaga91,frisch95}, {\it i.e.} the spatial inhomogeneity of the energy transfer across scales, that results in the generation of highly energetic, bursty small-scale fluctuations with scale-dependent, non-Gaussian statistics~\citep{sorriso99}. Such fluctuations are often referred to as coherent structures, with phase synchronization among a certain number of scales~\citep{greco14,lion16,roberts16,perrone16,perrone17}. At smaller scales, below the MHD range, a more variable level of intermittency is observed, whose nature is still under debate~\citep{AlexandrovaEA08,kiyani09}.   During the last years, the study of the intermittency and the consequent research on small-scale structures have been performed by using different methods. For example,   wavelet analysis~\citep{farge92,bruno99} provides the local estimation of the scale-dependent energy associated with the magnetic fluctuations, thus allowing the identification of the highly energetic intermittent structures. Alternatively, the Partial Variance of Increments (PVI) technique~\citep{greco08} was used to identify discontinuities, current sheets, and other intermittent features. 

%\item  Correlations between structures and heating etc.: Osman, Alexandros, Vlasov, Karimabadi, Thessein...
Recently, there is a compelling evidence that the kinetic processes are enhanced in the proximity of the turbulence-generated structures, which carry a larger amount of energy than the turbulent background. Measurements have shown that both ions and electrons are energized in the proximity of the most intense small-scale current sheets~\citep{osman11,osman12,tessein13,chasapis15,chasapis17}. Note that the same behavior has  been observed in HVM numerical simulations~\citep{servidio12,greco12,perrone14b}.
More recently, even reduced models such as giro-kinetics converged toward this general view \citep{Howes16,HowesEA18}. 
The processes responsible for such different forms of energization may involve magnetic reconnection~\citep{retino07,burch16}, wave-particle resonances, plasma instabilities~\citep{matteini13,servidio14,breuillard16} and enhancement of collisions~\citep{pezzi16a}. These topics are currently very debated in the space plasma community~\citep{chen16}.

%\item  Yaglom law in solar wind $\rightarrow$ Definition: local energy transfer rate proxy, LET, which terms?
In order to understand the interconnections between the turbulent structures and the small-scale kinetic processes, a more detailed description of the local energy flux across scales should be defined for plasma turbulence. This possible tool should take into account the complex nature of the nonlinear coupling between velocity and magnetic field. Recent studies included the contributions from cross-terms, typically associated to Alfv\'enic fluctuations, to better identify regions of small-scale accumulation of energy~\citep{sorriso2015}. 

In MHD, the energy transfer across scales obeys the Politano-Pouquet law (PP)~\citep{pp98}, which prescribes the linear scaling of the mixed third-order moment of the fields increments. The proportionality factor is the mean energy transfer rate $\langle\epsilon^\pm\rangle$, the brackets indicating ensemble average. For a two-dimensional field, the standard version of the PP law for the the mixed third-order moments $Y^\pm(l)$ is  
\begin{equation}
Y^\pm(l) = \langle |\Delta \bzpm_l(\br)|^2  \, \Delta z^\mp_{i,l}(\br) \rangle = 
-2  \langle \epsilon^\pm \rangle l \; .
\label{yaglom}
\end{equation}
$\Delta\psi_l=\psi(\br+l\br/r)-\psi(\br)$ is the longitudinal increment of a generic field $\psi$ (e.g. the magnetic field $\bB(\br)$) across the scale $l$; the subscript $i$ indicates the generic component of the field, the field increment being estimated along the same direction; $\bzpm=\bv\pm\bB/\sqrt{4\pi\rho}$ are the Elsasser variables that couple the plasma velocity $\bv$ and the magnetic field $\bB$, the latter being expressed in velocity units through the plasma density $\rho$. The validity of the PP law has been first verified in two-dimensional MHD numerical simulations~\citep{sorriso2002}, and later also in the solar wind~\citep{macbride2005,sorriso07,marino08,macbride2008,marino12}.

Based on the PP law~(Equation \ref{yaglom}), an heuristic proxy of the `local' energy transfer rate (LET) at the scale $l$ can be defined by introducing the quantity: 
\begin{equation}
\epsilon_l (\br) = -\frac{1}{2l}  \left[ |\Delta \bzp_l(\br)|^2  \, \Delta z^-_{i,l}(\br)  + |\Delta \bzm_l(\br)|^2  \, \Delta z^+_{i,l}(\br) \right] \, .
\label{pseudoenergy}
\end{equation}
At a given scale, each field increment in the time series can thus be associated with the local value of $\epsilon_l(\br)$~\citep{marschtu,sorriso2015,sorriso2018}. Note that the LET can be expressed in terms of velocity and magnetic field as $\epsilon_l(\br)\propto \Delta v_{i}(\Delta v^2+\Delta b^2)-2\Delta b_{i}(\Delta \mathbf{v}\cdot\Delta \mathbf{b})$. The first right-hand-side term is associated with the energy advected by the velocity fluctuations, and the second to the velocity-magnetic field correlations coupled to the longitudinal magnetic field fluctuations~\citep{fouad2017,sorriso2018}. Recently, by using Helios 2 data, it has been shown that the LET correctly identifies the regions of space in the turbulent solar wind that carry energy towards smaller scales. These regions are characterized by enhanced kinetic processes, resulting in higher proton temperature~\citep{sorriso2018}. It is worth noting that in recent works, the third order law in Eq.~(\ref{yaglom}) has been extended to include characteristic plasma length scales and compressibility~\citep{AndresEA18}.

The correlation between the local temperature increase and the large amplitude bursts of energy transfer suggests that the proxy defined in Eq.~({\ref{pseudoenergy}) might play an important role in the understanding of solar wind turbulent dissipation and heating. Moreover, the comparison between the properties of the PVI and LET has shown that both methods are suitable for the description of the turbulent energy cascade. Since the LET is a signed variable, in principle it should also carry information about the possible direction of the cross-scale energy flow, identifying possible local direct and inverse cascade processes. The latter cannot be studied by using PVI or wavelet-based techniques since these are positive-defined. 

The aim of this paper is to explore the link between the MHD turbulent energy cascade and the kinetic processes associated with deviations of the ion velocity distributions from the Maxwellian equilibrium. To this aim, we make use of the LET for interpreting high-resolution data from numerical simulations of the Vlasov-Maxwell multi-component system. The present work is organized as follows. In Section~\ref{sec:numres} the governing equations, the numerical algorithm and the parameters of the Vlasov simulations will be presented. A brief overview on the numerical results will be given. The use of the proxy of the local energy transfer rate will be the topic of Section~\ref{sec:heatres}, while in Section~\ref{sec:concls} we will discuss our conclusions.

\section{HVM multi-component simulations}
\label{sec:numres}

%\item  Vlasov: all the scales from fluid to kinetic (descrizione simulazioni e variabili usate 
%per diagnostica; spettro, intermittenza, multifrattalità, dissipazione...)
%check approx Yaglom in the MHD range of Vlasov (denise: check $<f(v)>_{dp}$) (stima di epsilon)
We use high resolution Vlasov simulations in order to simulate the turbulent multi-species plasma of the solar wind. First, high spatial resolution is needed in order to follow the evolution of the turbulent cascade up to wavelengths shorter than ion characteristic scales, where kinetic effects presumably govern the physics of the system. Second, high numerical resolution in the velocity domain is required to investigate the distortions of the ion distribution functions. The above two requirements are the main ingredients of the highly-performing, extensively tested, HVM code (http://fis.unical.it/hvm/).

\subsection{Numerical method}
We employ the HVM code which directly integrates the Vlasov equations for both proton and $\alpha$ particle distribution functions while electrons are considered as a massless isothermal fluid. The Vlasov equation for both ion species is combined with the Maxwell system, by assuming quasi-neutrality and by neglecting the displacement current. In particular Amp\'ere and Faraday equations and a generalized Ohm's law are solved. HVM equations, written in dimensionless form, are the following:
\begin{eqnarray}
 \frac{\partial f_s}{\partial t} + {\bf v} \cdot \frac{\partial f_s}{\partial {\bf r} } + 
 \xi_s \left( {\bf E} + {\bf v} \times {\bf B} \right) \cdot \frac{\partial f_s}{\partial {\bf v}} = 0 
 \label{eq:HVMvlas} \\
{\bf E} = - \left( {\bf u}_e \times {\bf B} \right) - \frac{\nabla P_e}{n_e} + \eta {\bf j} 
 \label{eq:HVMohm} \\
\frac{\partial {\bf B}}{\partial t} = - \nabla \times {\bf E} \;\;\; ; \;\;\; 
 \nabla \times {\bf B} = {\bf j} 
 \label{eq:HVMfar}
\end{eqnarray}
In previous equations, $f_s({\bf r}, {\bf v}, t)$ is the ion distribution function (the subscript $s=p,\alpha$ refers to protons and $\alpha$ particles, respectively), ${\bf E}({\bf r},t)$ and ${\bf B}({\bf r},t)$ are the electric and magnetic fields, respectively, and $\xi_s$ is a constant reflecting the charge to mass ratio of each species ($\xi_p=1$, $\xi_\alpha=1/2$). The electron bulk speed is ${\bf u}_e = \left( \sum_s Z_s n_s {\bf u}_s - {\bf j} \right)/n_e$, where $Z_s$ is the ion charge number ($Z_p=1$, $Z_\alpha=2$), the ion density $n_s$ and the ion bulk velocity ${\bf u}_s$ are the zero-th and first order moments of the ion distribution function, respectively, and the electron density $n_e$ is evaluated through the quasi-neutrality condition $n_e= \sum_s Z_s n_s$. Furthermore, $P_e=n_e T_e$ is the electron pressure ($T_e$ is homogeneous and stationary) and ${\bf j}=\nabla \times {\bf B}$ is the total current density. The external small resistivity $\eta=2\times 10^{-2}$ is introduced to remove spurious numerical instabilities. In Equations~(\ref{eq:HVMvlas})--(\ref{eq:HVMfar}), masses, charges, time, velocities and lengths are respectively normalized to the proton mass, $m_p$, and charge, $e$, to the inverse of the proton cyclotron frequency, $\Omega_{cp}^{-1} = m_p c / e B_0$ (being $c$ and $B_0$ the speed of light and the ambient magnetic field), to the Alfv\'en velocity, $V_A = B_0 / \sqrt{4\pi n_{0,p} m_p}$ (being $n_{0,p}$ the equilibrium proton density), and to the proton skin depth, $d_p= V_A/ \Omega_{cp}$. 

The described set of Equations~(\ref{eq:HVMvlas})--(\ref{eq:HVMfar}) are solved in a $2.5D-3V$ phase space domain, where $2.5D$ means that all vector components are retained but they depend only on the two spatial coordinates $(x,y)$. The spatial domain, whose size is $L=2\pi\times 20 d_p$ in both directions, has been discretized with $512^2$ grid-points and, at boundaries, periodic conditions are implemented. Each direction of the velocity domain of each species (protons and $\alpha$ particles) is discretized with $71$ grid-points in the range $v_{j,s}=\left[-v_{max,s},v_{max,s}\right] \;\; (j=x,y,z)$, where $v_{max,s}=\pm 5 v_{th,s}$ (being $v_{th,s}=\sqrt{k_B T_{0,s}/m_s}$ the ion thermal speed, with $T_{0,s}$ the equilibrium temperature). Boundary conditions in velocity space assume $f_s(|v| > v_{max,s})= 0$ in each direction. 
The time step $\Delta t$ has been chosen in such a way that the CFL condition for the numerical stability of the Vlasov algorithm is satisfied~\citep{peyret86}. To control the numerical accuracy of the simulation, the basic set of conservation laws of Vlasov equation is continuously monitored during the simulations: relative variation of proton mass, $\alpha$ particle mass, total energy and entropy are about $2.5\times 10^{-3}\%$, $3.5\times 10^{-1}\%$, $0.6\%$ and $0.4\%$, respectively.

At the initial time of the simulation ($t=0$), both ion species have Maxwellian distribution functions, with homogeneous densities, and an uniform background out-of-plane magnetic field ${\bf B_0} = B_0 {\bf e}_z$ is imposed. In order to mimic the physical conditions of the pristine solar wind, we assume that i) the $\alpha$ particle to proton density ratio $n_{0,\alpha}/n_{0,p}=5\%$ and ii) we consider $\alpha$ particles, protons and electrons at the same initial temperature $T_{0,\alpha}=T_{0,p}=T_{e}$. The equilibrium is then perturbed through a $2D$ spectrum of Fourier modes in the plane $(x,y)$, for both proton bulk velocity and magnetic field, such that proton bulk velocity and magnetic field are globally uncorrelated. Energy is injected with wave numbers in the range $0.1 < k < 0.3$ ($2 \le m \le 6$, where $k=2\pi m/L$) and random phases. We did not introduce neither density disturbances nor parallel variances, i.e. $\delta n = \delta u_z = \delta B_z = 0$ at $t=0$. The r.m.s. level of fluctuations is $\delta B/B_0 = 1/3$, while $\beta_p= 2 v_{th,p}^2/ V_A^2 = 0.5$.

The kinetic evolution of protons and $\alpha$ particles is numerically investigated in a decaying turbulence. During the simulation, nonlinear couplings between fluctuations generate a turbulent cascade, where the injected energy is transferred along the spectrum to smaller, kinetic scales. The time $t^*$ of maximum  turbulent activity can be identified as the peak time of the spatially averaged out-of-plane current density, $\langle j_z^2\rangle$, a standard indicator for small-scale magnetic gradients. All the analysis has been performed using one snapshot of the simulation at $t^* \Omega_{cp} = 49$, when decaying turbulence approaches a quasi-steady state. 
A more detailed description of the numerical results of the simulation can be found in~\cite{valentini16}.

\subsection{Numerical results}

An overview of the differential kinetic dynamics of the ion species, namely protons and $\alpha$ particles, associated to the development of the turbulent cascade, is presented in Figure~\ref{fig:maps}. 
%%%%%%%%%%%%%%%%%%%%%%%%%%%%%%%%%%%%%%%%%%%%%%%
%	FIG 1: fields 2D   
%%%%%%%%%%%%%%%%%%%%%%%%%%%%%%%%%%%%%%%%%%%%%%%
\begin{figure*}    
\centering
\epsfxsize=\textwidth \centerline{\epsffile{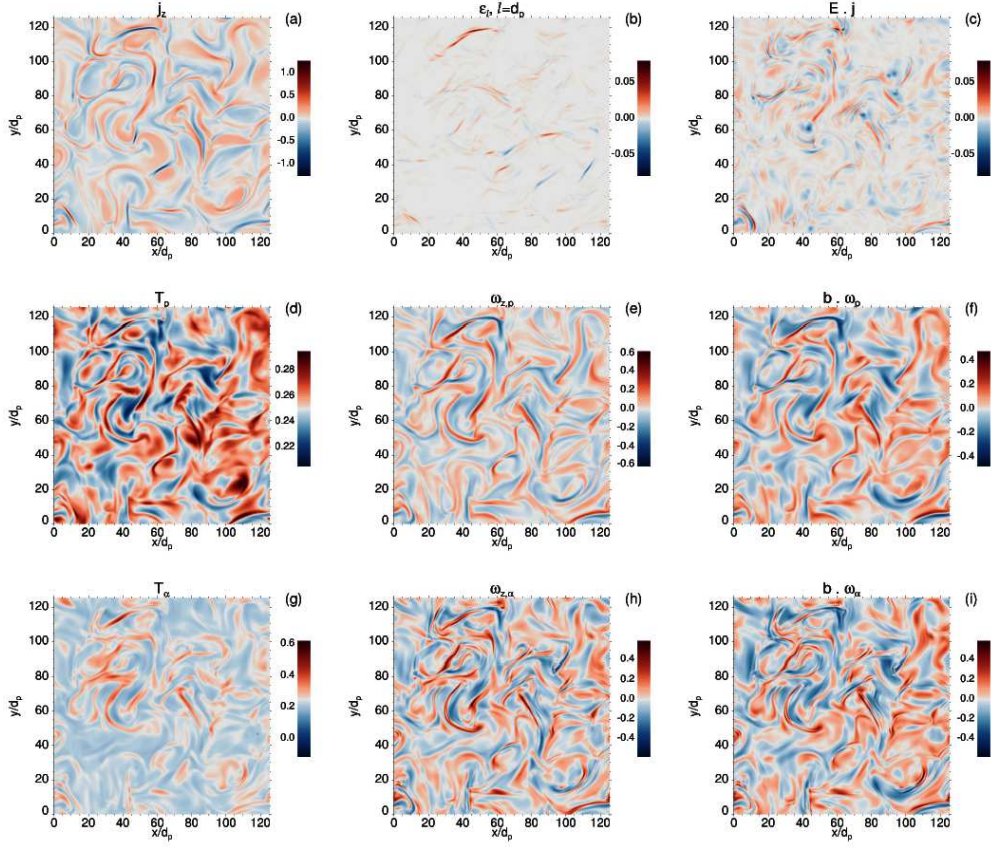}}   
\caption{(Color online) 2D contour plots for the HVM multi-component simulation at $t=t^*$. (a) The out-of-plane total current density $j_z$; (b) the local energy transfer rate $\epsilon_{d_p}$; (c) ${\bf E} \cdot {\bf j}$; (d) the proton temperature $T_p$; (e) the out-of-plane proton vorticity $\omega_{z,p}$; (f) ${\bf b} \cdot \omega_{p}$; (g) the $\alpha$ particle temperature $T_{\alpha}$; (h) the out-of-plane $\alpha$ particle vorticity $\omega_{z,\alpha}$; and (i) ${\bf b} \cdot \omega_{\alpha}$.}
\label{fig:maps}
\end{figure*}
%%%%%%%%%%%%%%%%%%%%%%%%%%%%%%%%%%%%%%%%%%%%%%%
Turbulence leads to the generation of small-scale coherent structures, such as current sheets, filaments, or strongly shared flows, which can be recognized in the two-dimensional patterns in physical space. In particular, panel (a) of Figure~\ref{fig:maps} shows the shaded contours of the out-of-plane current density, $j_z$.  The current density becomes very intense in between magnetic islands, where magnetic reconnection can locally occur. 

Panel (b) shows the contour map of the local energy transfer rate, $\epsilon_l (x,y)$, estimated using the velocity and magnetic field (in Alfv\'en units) two-point increments at the scale $l=d_p$. Positive and negative filaments of strong energy transfer $\epsilon_l$ are present, but the spatial correlation with strong current sheets is unclear. There is a shift between the current layers and this cascade-transfer indicator, probably due to the fact that the former are the sites of dissipation, while the LET is high at the very active nonlinear regions, such as the upstream parts of the reconnecting current sheets. 

In panel (c) we display the contour map of ${\bf E} \cdot {\bf j}$, a proxy of energy conversion between fields and particles~\citep{wan2015,yang2017}. Large positive values are observed in the proximity of the strongest current sheets, suggesting that the electromagnetic energy might be converted into kinetic and thermal energy where magnetic reconnection is likely to occur. 

Panels (d) and (g) of Figure~\ref{fig:maps} show the maps of proton and $\alpha$-particle temperature, respectively. Comparison with panel (a) highlights that the presence of strong current sheets corresponds to temperature enhancement of both ions. Although the initial conditions of the simulation infer equal and uniform temperature for the two ion species, during the turbulent evolution of the system both ions show a non-uniform temperature pattern, with similar inhomogeneity as for the current $j_z$. Color bars were chosen so that the initial temperature $T_0$ is white, while red and blue indicate local heating and cooling, respectively. A definite temperature increase can be observed close to the region of strong magnetic field gradients, meaning that the heating processes are influenced by the topology of the magnetic field. Preferential heating is also evident, as the $\alpha$ particles temperature is nearly twice that of the protons. Moreover, while the protons display roughly equal heating and cooling with respect to the initial temperature $T_0$, the $\alpha$ particles show predominant temperature increase. 

Enhanced ion heating was recently observed in spatial regions nearby the peaks of the ion vorticity $\omega = \nabla \times {\bf v}$~\citep{servidio15,franci16,valentini16}.  Previous studies have shown the association between viscous-like effects, velocity shears~\citep{markovskii06,delsarto16} and vorticity~\citep{huba96}, including the dependence on the sign of ${\bf B_0} \cdot {\bf \omega}$~\citep{parashar16}.  In particular, it has been shown that the vorticity may enhance local kinetic effects, with a generalized resonance sign-dependence condition for the enhancement or reduction of the ion heating and temperature anisotropy~\citep{parashar16}. In order to explore such correlation, the maps of the out-of-plane ion vorticities $\omega_{z,p}$ and $\omega_{z,\alpha}$, are shown in panels (e) and (h) of Figure~\ref{fig:maps}. A pattern of small-scale vorticity structures can be observed, similar to the case of the current. However, a more complex topology is observed for the $\alpha$ particles vorticity, with the splitting of the structures in smaller filaments already observed in Hall-MHD turbulence~\citep{martin}. 

Finally, in panels (f) and (i) of Figure~\ref{fig:maps} we show the 2D contour plot of $ {\bf b} \cdot {\bf \omega}$ for both protons and $\alpha$ particles, respectively, where ${\bf b} = {\bf B} - {\bf B_0}$ indicates the magnetic field fluctuations. A behavior similar to $\omega_z$ is observed, so that both positive $\omega_z$ and $ {\bf b} \cdot {\bf \omega}$ correspond to increased temperature, for both ion species. 

%%%%%%%%%%%%%%%%%%%%%%%%%%%%%%%%%%%%%%%%%%%%%%%
%	FIG 2:  Correlations
%%%%%%%%%%%%%%%%%%%%%%%%%%%%%%%%%%%%%%%%%%%%%%%
 \begin{figure*}    
 \centering
  \begin{minipage}{0.45 \textwidth}
    \epsfxsize=6.5cm \centerline{\epsffile{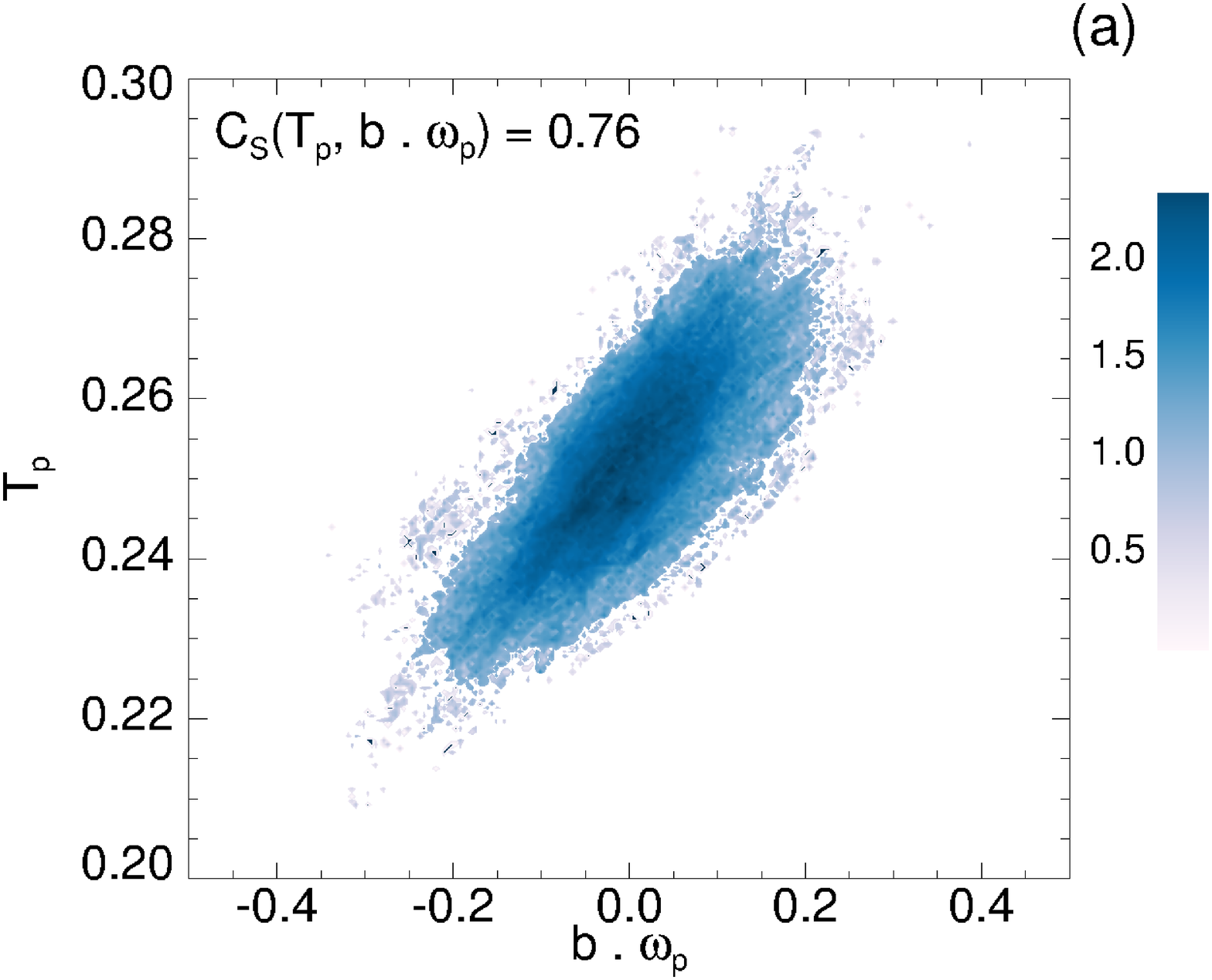}}   
  \end{minipage}
 \hfill
  \begin{minipage}{0.45 \textwidth}
    \epsfxsize=6.5cm \centerline{\epsffile{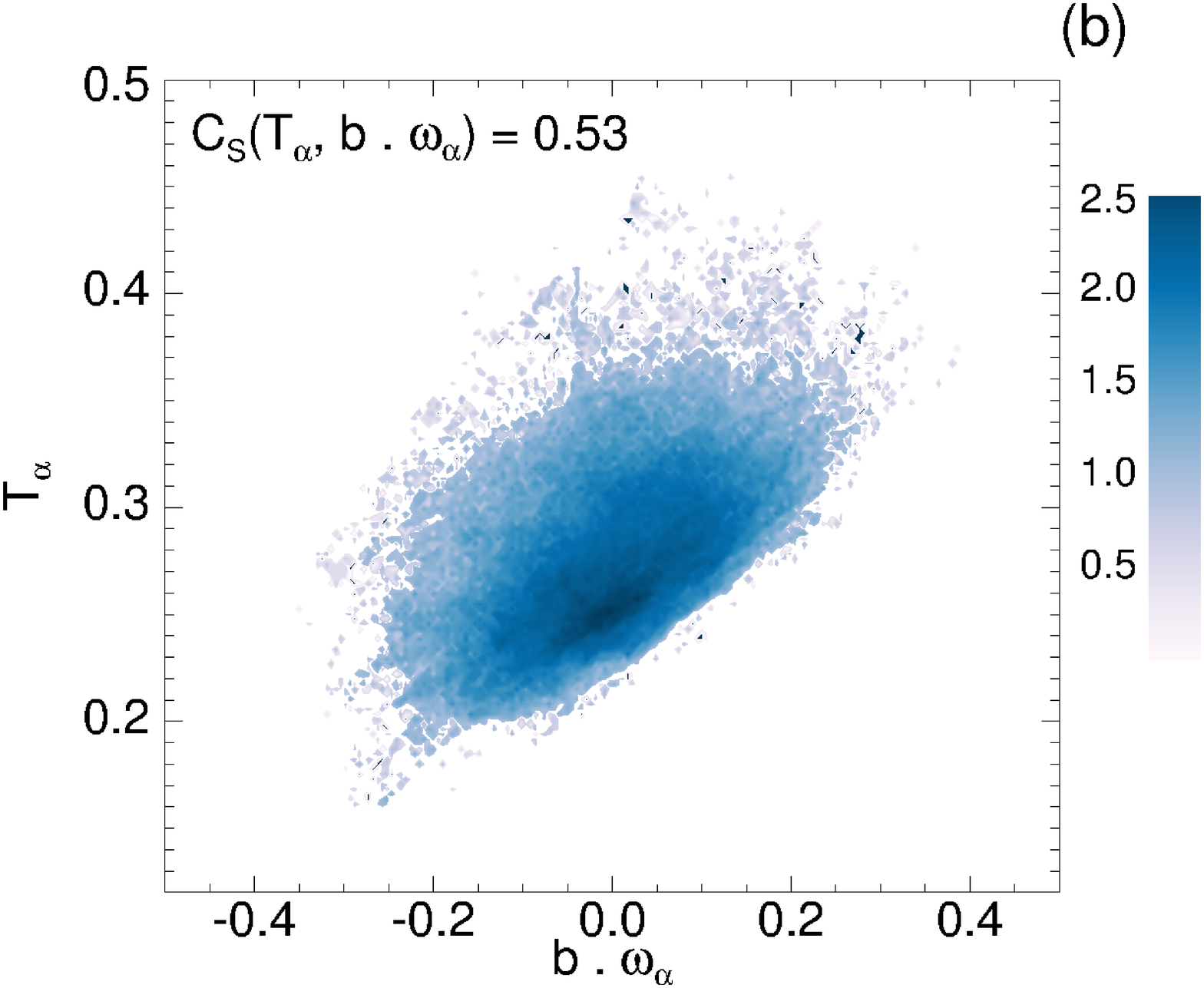}}   
  \end{minipage}
 \hfill 
  \begin{minipage}{0.45 \textwidth}
    \epsfxsize=6.5cm \centerline{\epsffile{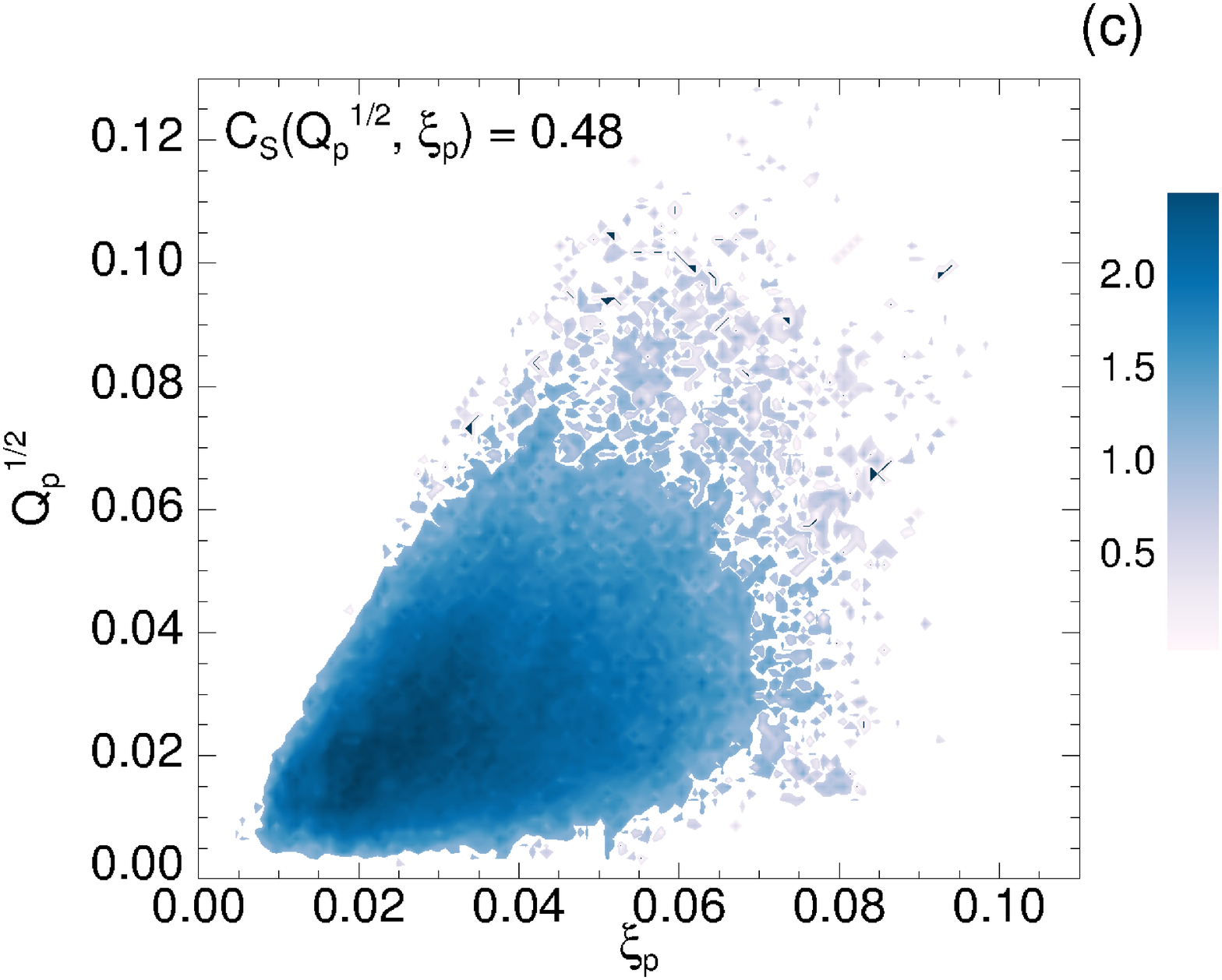}}   
  \end{minipage}
 \hfill
  \begin{minipage}{0.45 \textwidth}
    \epsfxsize=6.5cm \centerline{\epsffile{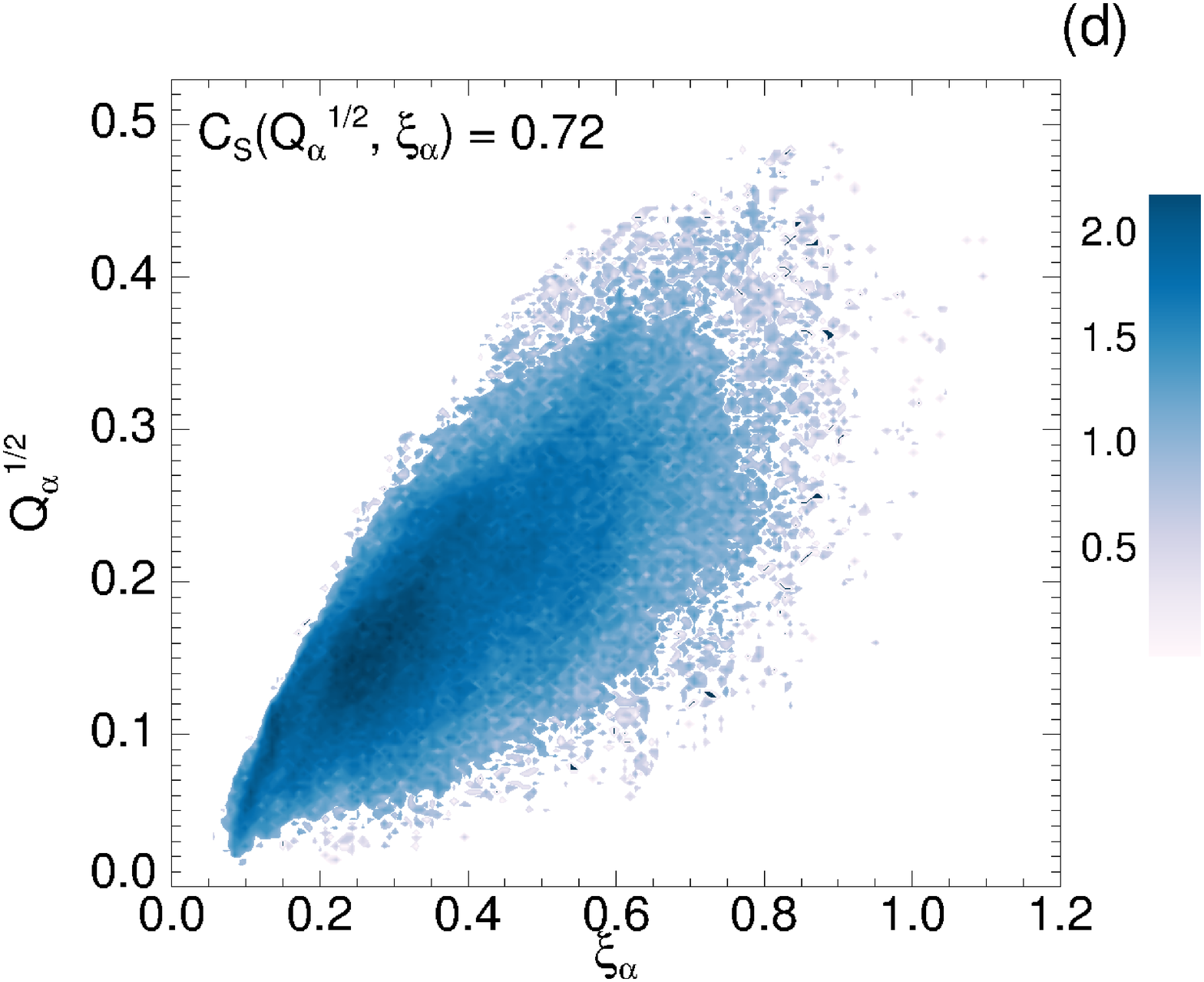}}   
  \end{minipage}
  \caption{Joint probability distributions for (a) $T_p$ and ${\bf b} \cdot  {\bf \omega}_p$; (b) $T_\alpha$ and ${\bf b} \cdot  {\bf \omega}_{\alpha}$; (c) $Q_p^{1/2}$ and $\xi_p$; (d) $Q_{\alpha}^{1/2}$ and $\xi_\alpha$.}
  \label{fig:correlation}
 \end{figure*}
%%%%%%%%%%%%%%%%%%%%%%%%%%%%%%%%%%%%%%%%%%%%%%%
The overview of the numerical results of the HVM multi-species simulation given in Figure~\ref{fig:maps} highlights a general enhancement of the ion temperature related to the presence of thin current sheets. A more evident correlation is found with the ion out-of-plane vorticity $\omega_s$ and with ${\bf b} \cdot {\omega}_s$. To better appreciate the robustness of the observed correlation between $T_s$ and ${\bf b} \cdot {\bf \omega}_s$, in panel (a) e (b) of Figure~\ref{fig:correlation} we show the average joint probability distributions $P(T_p, {\bf b} \cdot  {\bf \omega}_p)$ and $P(T_{\alpha}, {\bf b} \cdot {\bf \omega}_{\alpha})$, respectively. A good correlation is observed for both ion species, that we quantify through the nonlinear Spearman correlation coefficients, $C_S$, indicated in each panel of Figure~\ref{fig:correlation}. Correlations are stronger for protons $[C_S(T_p, {\bf b} \cdot {\bf \omega}_p)=0.76]$, than for $\alpha$ particles $[C_S(T_{\alpha}, {\bf b} \cdot {\bf \omega}_{\alpha})=0.53]$.  However, the correlation between the temperature and the out-of-plane vorticity ({\it i.e.}, the vorticity component parallel to the background magnetic field rather than to the magnetic fluctuations) is larger for the $\alpha$ particles: $C_S(T_p, \omega_{z,p})=0.49$ and $C_S(T_{\alpha}, \omega_{z,\alpha})=0.63$ (joint distributions are not shown for this case).
The heating is thus better correlated with the out-of-plane vorticity for the $\alpha$ particles, but with the vorticity component parallel to the local magnetic fluctuations for the protons. This may be related to the heavier $\alpha$ particles being slower than protons in following the magnetic field fluctuations. 

The heating mechanisms are also correlated with strong deformation of the ion velocity distribution~\citep{servidio12,valentini16}. For each ion species and in each point of the simulation domain, the non-Maxwellian measure can be defined as~\citep{greco12, vasconez15, pucci16, valentini17, pezzi17c}:
\begin{eqnarray}
 \xi_s = \frac{1}{n_s}\sqrt{\int{(f_s-g_s)^2 d^3v}} .
 \label{eq:epsilonSergio}
\end{eqnarray}
The above scalar function measures the deviation of the observed distribution $f_s$ from the associated Maxwellian $g_s$, which is built using the local density, bulk velocity and temperature. The latter quantity is also intimately related to the so-called velocity space cascade, recently observed in space plasmas~\citep{servidio17}.

It is useful at this point to considered the agyrotropy of the pressure tensor, {\it i.e.} its departure from cylindrical symmetry around the local magnetic field direction. Non-gyrotropy is important as a known source of instabilities, and may help identifying spatial  structures in the plasma~\citep{scudder08,aunai13,swisdak16}.  For both ion species, given the 3{$\times$}3 pressure tensor with elements $P_{ij}$ in an arbitrary Cartesian system with coordinates $(x,y,z)$, the measure of agyrotropy can be defined as the scalar~\citep{swisdak16}:
\begin{eqnarray}
Q_s = 1 - \frac{4 I_{2,s}}{(I_{1,s}-P_{\parallel,s})(I_{1,s}+3P_{\parallel,s})} \ ,
\end{eqnarray}
with $I_{1}=P_{xx}+P_{yy}+P_{zz}$, and $I_{2}=P_{xx}P_{yy}+P_{xx}P_{zz}+P_{yy}P_{zz}-(P_{xy}P_{yx}+P_{xz}P_{zx}+P_{yz}P_{zy})$. $P_{\parallel}$ here represents the parallel pressure, where we have omitted the subscript $s$ for simplicity.  A good correlation is observed between the non-gyrotropy of the ion pressure tensor, $Q_s^{1/2}$, and the non-Maxwellianity of the ion velocity, $\xi_s$, as shown in the joint probability distribution maps in panels (c) and (d) of Figure~\ref{fig:correlation} for protons and $\alpha$ particles, respectively. This mechanism works better for $\alpha$ particles $[C_S(Q_\alpha^{1/2}, \xi_\alpha)=0.72]$ with respect to protons $[C_S(Q_p^{1/2}, \xi_p)=0.48]$, since deviations from Maxwellian are significantly stronger for the $\alpha$ particles than for protons. 

Table~\ref{tab:spearman} collects the Spearman correlation coefficients estimated for pairs of quantities for which $C_S \gtrsim 0.4$, namely if at least a significant moderate correlation is present. These also include the temperature anisotropy $A_s=T_{\perp,s}/T_{\parallel,s}$, defined for both ion species, an important diagnostic for kinetic processes and instabilities. In the Table, we have omitted the weakly correlated pairs, for which $C_S\lesssim 0.4$. 
%
% TABLE
%
\begin{table*}
  \begin{center}
\def~{\hphantom{0}}
  \begin{tabular}{ccccc}

\hline
\multicolumn{5}{c}{Spearman Correlation Coefficients} \\%[1pt]
\hline
 $(T_p, \omega_{z,p})=0.49$ & \null & $(T_p, {\bf b} \cdot {\bf \omega}_p)=0.76$ & \null & $(T_p, A_p)=0.42$ \\
\hline 
 $(T_\alpha, \omega_{z,p})=0.63$ & \null & $(T_\alpha, {\bf b} \cdot {\bf \omega}_\alpha)=0.53$ & \null & 
 $(T_\alpha, A_\alpha)=0.52$ \\
\hline 
 $(A_p, \omega_{z,p})=0.45$ & \null & $(A_p, {\bf b} \cdot {\bf \omega}_p)=0.38$ & \null & $(Q_p^{1/2}, \xi_p)=0.48$ \\
\hline
 $(A_\alpha, \omega_{z,\alpha})=0.57$ & \null &
 $(A_\alpha, {\bf b} \cdot {\bf \omega}_\alpha)=0.52$ & \null & $(Q_\alpha^{1/2}, \xi_\alpha)=0.72$ \\
\hline 
 $(\omega_{z,\alpha}, \omega_{z,p})=0.65$ & \null & $({\bf b} \cdot {\bf \omega}_\alpha, {\bf b} \cdot {\bf \omega}_p)= 0.69$ & \null &
 $(\xi_{z,\alpha}, \xi_{z,p})=0.61$ \\
 \hline
  \end{tabular}
  \caption{Nonlinear Spearman correlation coefficients, $C_S$, between pairs of quantities for which $C_S\gtrsim 0.4$. All quantities are estimated using the HMV simulation results at the time $t=t^*$.}
  \label{tab:spearman}
  \end{center}
\end{table*}

To summarize the results of the above analysis, larger values of the deformation of the ion velocity and of the pressure tensor agirotropy are recovered in spatial positions nearby the peaks of the ion vorticity and close to peaks of the current density. In these regions, differential heating and temperature anisotropy are also observed. Ion heating and kinetic effects are in general more strongly correlated with the vorticity than with the current density profile.

\section{Local energy transfer rate in Vlasov-Maxwell simulations}
\label{sec:heatres}

% \item Yaglom
\subsection{Properties of the local energy transfer rate}
Incompressible and isotropic MHD turbulence in plasmas can be described by the exact relation for the energy flux through the scales (see Equation~\ref{yaglom}). This Yaglom-like scaling law has been extensively tested in the solar wind, recovering very good agreements. The use of HVM multi-component simulation, described in Section~\ref{sec:numres}, can provide a better understanding of the connection between the MHD turbulent cascade and the relative kinetic effects on the ion velocity distributions since the `in situ' solar-wind measurements have several limitations. 

In Figure~\ref{fig:yaglom}, we present the validation of the PP law for the turbulent Vlasov simulation used in this work, as estimated by averaging Equation~(\ref{pseudoenergy}). 
In the range of scales roughly corresponding to the inertial range of MHD turbulence, {\it i.e.} for $2d_p \lesssim l \lesssim 10d_p$, the scaling of $-Y(l)$ is compatible with a linear law (Figure~\ref{fig:yaglom}). This suggests that the turbulent energy cascade is well developed in such range, at the time $t^*$ of the simulation. However, the limited extension of the inertial range does not allow a more precise and quantitative estimate \citep{parashar15}.
It should be pointed out that the law as defined in Equation~(\ref{yaglom}), without absolute value, is indicating the direction of the cascade towards small scales. The limited extension of the range observed in the HVM simulation may arise from several facts, which are physically meaningful. First, the system described here is size-limited due to computational constraints: the correlation length is limited to roughly 10$d_p$ ---relatively short as compared to the several decades of MHD turbulence commonly observed in the solar wind. Second, the PP law is exact only under the assumption of full isotropy, while our 2D simulation has a mean magnetic field. Third, the ion skin depth is a characteristic length of the system, which breaks the validity of the MHD law. A more precise description could be obtained in terms of the Hall MHD version of the PP law~\citep{AndresEA18}. This development is left for a future work. Fourth, in the derivation of these MHD-based scaling laws, isotropy of the pressure tensor is assumed, which does not hold  in Vlasov turbulence. Finally, processes such as ion-cyclotron resonances and Landau damping might transfer free energy from the physical space to the velocity space, producing non-Maxwellian effects and, eventually, heating. All these caveats suggest that the analysis in terms of PP law and LET can only be meaningful for increments in the inertial range, $l>d_p$.
%%%%%%%%%%%%%%%%%%%%%%%%%%%%%%%%%%%%%%%%%%%%%%%
%	FIG 3:  Yaglom
%%%%%%%%%%%%%%%%%%%%%%%%%%%%%%%%%%%%%%%%%%%%%%%
\begin{figure}
\centering
\epsfxsize=0.8\textwidth \centerline{\epsffile{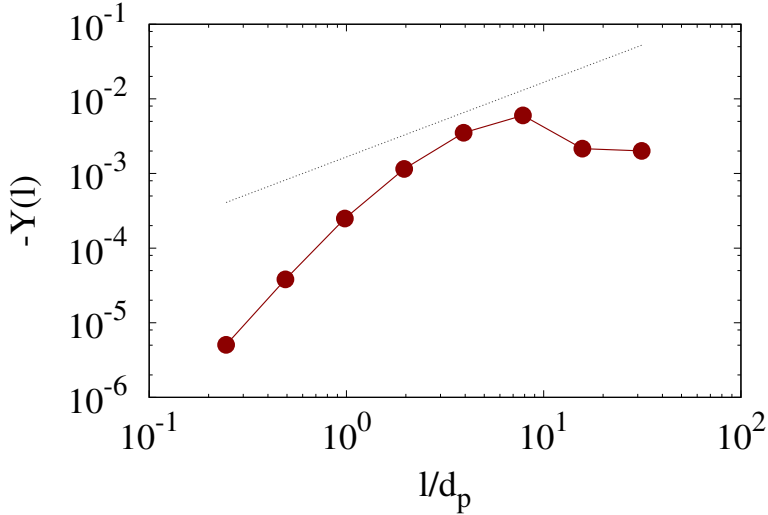}}
\caption{(Color online) The Yaglom law validation for turbulent Vlasov simulation (red markers and dashed line). The averaged scaling $Y = (Y^+ + Y^-)/2$ is shown.
The black dotted line represents an example of linear scaling, and is shown for reference only.}
\label{fig:yaglom}
\end{figure}
%%%%%%%%%%%%%%%%%%%%%%%%%%%%%%%%%%%%%%%%%%%%%%%

The local proxy $\epsilon_l(x,y)$ has been evaluated at different spatial scales $l$ using the HVM simulation results. 
%\item scalogrammi
In Figure~\ref{fig:scalogram} we show the scalogram of the LET $\epsilon_l$ along a one-dimensional cut of the 2D spatial domain of the simulation. Note that a similar analysis has been performed for the PVI scalograms~\citep{GrecoEA16}. Here we consider a diagonal path $r$, normalized to $d_p$, that crosses the simulation box several times. The channels of energy being transferred across scales are clearly visible, and reveal the complex, interwoven pattern of positive and negative flux that determines the neat energy flux~\citep{coburn14,sorriso2018}. 
%
%%%%%%%%%%%%%%%%%%%%%%%%%%%%%%%%%%%%%%%%%%%%%%%
%	FIG 5:  scalogram of epsilon
%%%%%%%%%%%%%%%%%%%%%%%%%%%%%%%%%%%%%%%%%%%%%%%
\begin{figure}
    \centering
 \epsfxsize=0.9\textwidth \centerline{\epsffile{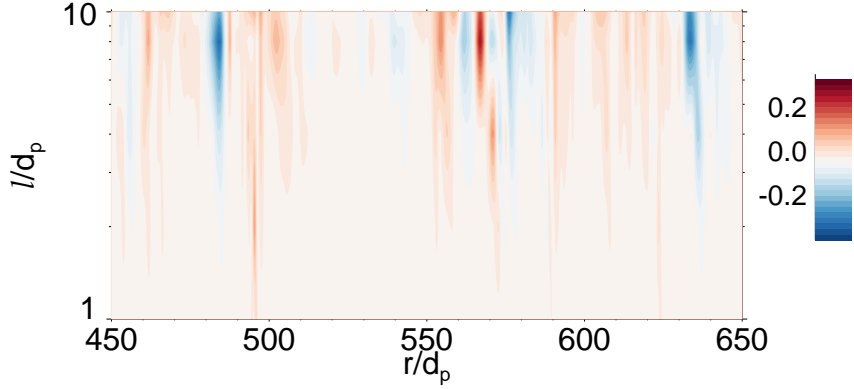}}
 \caption{(Color online) The scalogram of LET $\epsilon_l$, plotted as color map at different scales $l/d_p$, estimated at different positions $r/d_p$ along a one-dimensional cut of the simulation domain. The alternate positive and negative energy transfer regions are evident.}
 \label{fig:scalogram}
\end{figure}
%%%%%%%%%%%%%%%%%%%%%%%%%%%%%%%%%%%%%%%%%%%%%%%

% PDF
In order to characterize the statistical properties of the local energy transfer rate, in Figure~\ref{fig:pdfEps}, we show the probability distribution functions (PDFs) of the negative (left panel) and positive (central panel) standardized LET values ({\it i.e.}, $\epsilon_l(x,y)$ has been rescaled to have zero mean and unit standard deviation), for three different scales, namely $l=d_p$ (blue diamonds), $4d_p$ (green triangles) and $8d_p$ (red circles). In agreement with the PP law, for a given scale the positive tails are considerably higher and shallower than the negative ones, indicating the neat energy flux corresponding to the direct turbulent energy cascade towards small scales. This is clearly visible from the PDFs represented in Figure~\ref{fig:pdfEps}. 

Both the positive and negative PDFs behave roughly as stretched exponential functions $\sim \exp{(-b\epsilon_l^c)}$, where $c$ is the parameter that controls the shape of the distribution tails. In particular, $c=2$ indicates a Gaussian tail, $c=1$ an exponential tail, and for $c\to 0$ the tail approaches a power-law. The curvature parameter $c$, obtained from the fit at each scale, is shown in the right panel of Figure~\ref{fig:pdfEps}. Starting from the exponential value $c\sim 1$ at large scales (say for $l\gtrsim 10d_p$), its value decreases with the scales, indicating the formation of highly localized small-scale structures, typical of intermittency. The two tails of the PDFs have different scaling, so that the positive tail has a more evident, faster decay than the negative tail. Such difference quantitatively confirms the imbalance already observed through the PDF shape.
In the range corresponding to the MHD inertial range, the scaling of $c$ for the positive tail is well described by a power-law $c^\gamma_{\pm}$, with exponent $\gamma_{+}\sim 0.92\pm0.06$. For the negative tail, the scaling is less evident, but can be considered compatible with a power law with exponent $\gamma_{-}\sim 0.37\pm 0.13$. This confirms that the numerical simulation has reached a stage of fully developed turbulence, with a strongly intermittent energy cascade, whose characteristics are well captured by the LET. Note that a similar behavior has been recently observed in solar wind measurements~\citep{sorriso2018}. 
At smaller scales, for $l\lesssim d_p$, the parameter seems to reach a limiting value for both the positive and negative tails, indicating the saturation of intermittency. 
In this short range of scales, the positive tail of the PDF is described by $c\sim 0.1$, closer to power-law decay. The negative tail converges towards a slightly larger value $c\sim 0.3$, indicating a faster than power-law decay, and confirming once more the quantitative difference between positive and negative energy transfer rate.

%%%%%%%%%%%%%%%%%%%%%%%%%%%%%%%%%%%%%%%%%%%%%%%
%	FIG 4:  PDF epsilon
%%%%%%%%%%%%%%%%%%%%%%%%%%%%%%%%%%%%%%%%%%%%%%%
\begin{figure}
\centering
\epsfxsize=\textwidth \centerline{\epsffile{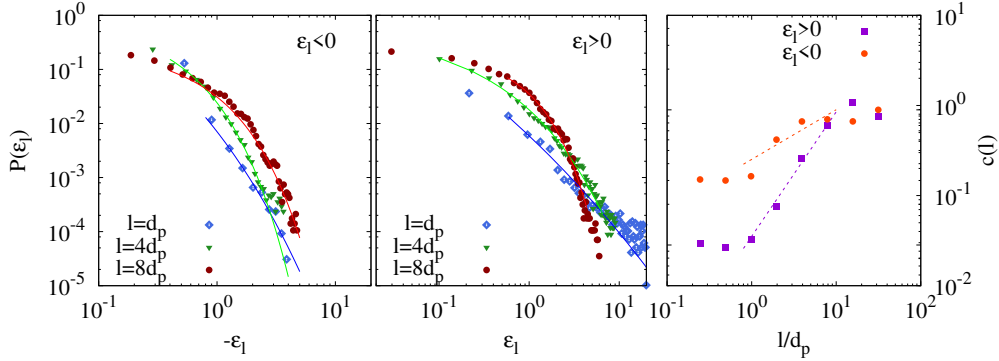}}
\caption{(Color online) In the left panel, the probability distribution functions (PDFs) of the standardized, negative $\epsilon_l(x,y)$ are shown for three different scales, namely $l=d_p$ (blue diamonds), $4d_p$ (green triangles) and $8d_p$ (red circles), using logarithmic axes. In the central panel, the PDFs of positive $\epsilon_l(x,y)$ are shown for the same scales, with the same symbols. In both panels, the different lines are the stretched exponential fits, obtained through a standard least-$\chi^2$ procedure. The right panel shows the scaling properties of the curvature parameter $c(l)$, for the positive (violet squares) and negative (orange circles) tails of the PDFs. The inertial range power-law fits are also indicated as dashed lines, the scaling exponents being respectively $\gamma_{+}=0.92$ and $\gamma_{-}=0.37$.}
\label{fig:pdfEps}
\end{figure}
%%%%%%%%%%%%%%%%%%%%%%%%%%%%%%%%%%%%%%%%%%%%%%%

%Osman
\subsection{Conditioned analysis of kinetic quantities}
After establishing the statistical properties of the LET in our hybrid simulation, we have performed the conditioned average analysis proposed by~\citet{osman12}, and recently performed using the LET~\citep{sorriso2018}. The conditioned analysis is designed to highlight the existence of robust features in the data that might be hidden by the turbulent fluctuations. To this goal, we have selected different values of threshold $\theta_{d_p}$ of the LET $\epsilon_{d_p}$, estimated at the bottom of the inertial range $l\simeq 2d_p$, to identify locations associated with large energy flux at the end of the fluid-like nonlinear turbulent cascade. Then, for each threshold, some of the variables described in Section~\ref{sec:numres} have been averaged only in those locations where the LET overcomes the threshold: $\epsilon_{d_p} \geq \theta_{d_p}$ for positive thresholds (cascade towards smaller scales), and  $\epsilon_{d_p} \leq \theta_{d_p}$ for negative thresholds (cascade towards higher scales). This gives the conditioned averaged values, for example $\langle T_p | \theta_{d_p} \rangle$ in the case of the proton temperature.

Through the above procedure, the turbulent fluctuations are all averaged to their mean value, except where systematic variations are present. The same procedure is repeated using all the points located at a given distances $\delta$ from the location of large LET, and for different values of $\delta$. This provides the average profile of the physical quantities in the vicinity of the bursts of energy flux, as a function of the distance from the LET burst. By repeating the same analysis using different thresholds, it is then possible to determine the role of the energy flux in the robust modification of the physical quantities. 

%%%%%%%%%%%%%%%%%%%%%%%%%%%%%%%%%%%%%%%%%%%%%%%
%	FIG 6:  Osman
%%%%%%%%%%%%%%%%%%%%%%%%%%%%%%%%%%%%%%%%%%%%%%%
\begin{figure}  
\centering
\epsfxsize=\textwidth \centerline{\epsffile{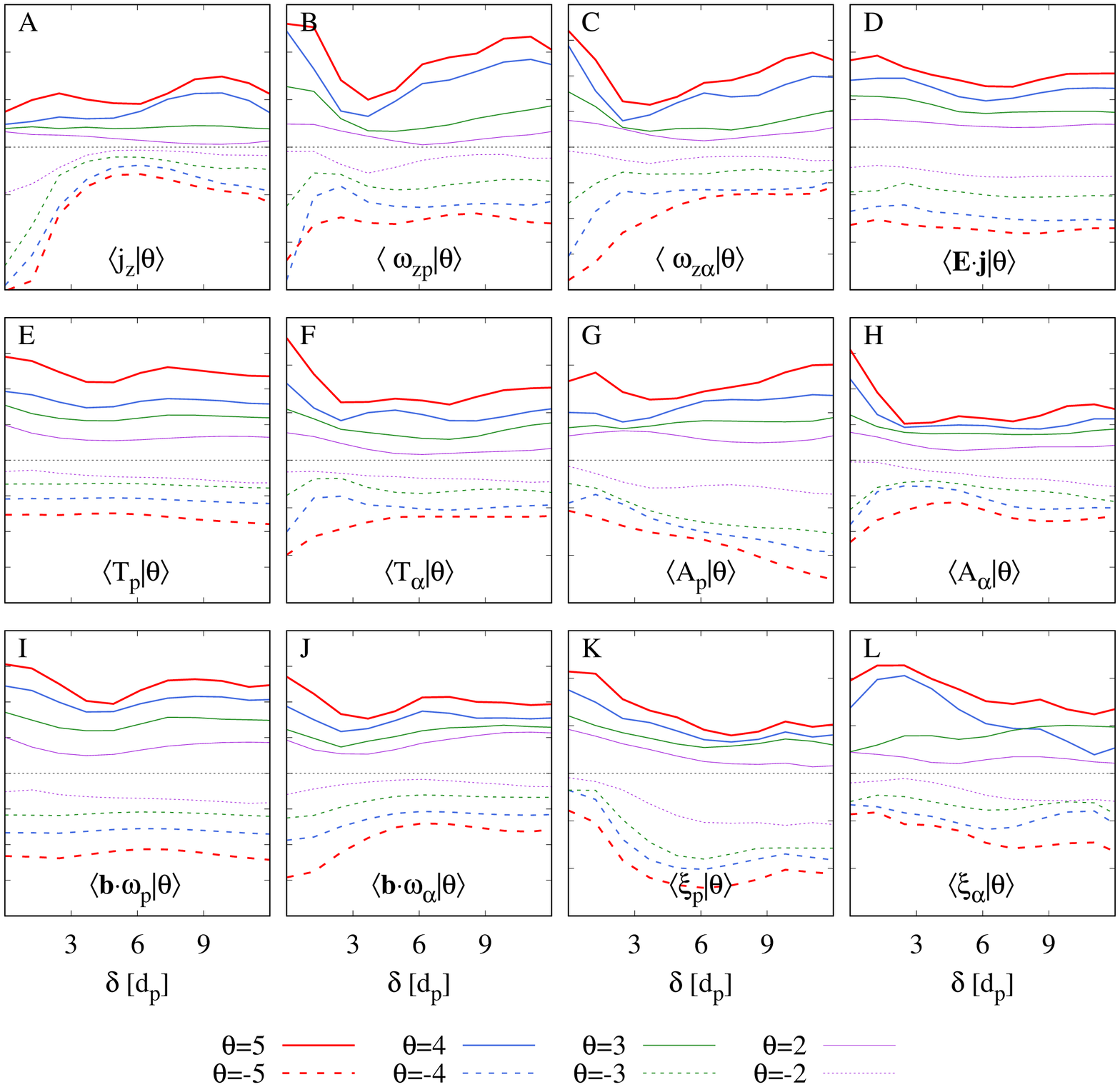}}    
\caption{(Color online) Conditioned averaged quantities as a function of the distance from the peak of the LET, estimated at the scale $l\simeq 2d_p$, for four positive and four negative LET threshold values $\theta_{d_p}$ (coded in different line colors, as indicated at the bottom of the Figure). Each panel refers to one of the physical quantities (see Section~\ref{sec:numres}, given in the legend located at the bottom of the panel). All curves have been vertically shifted for clarity, and are thus presented in arbitrary units. In each panel, the curves for larger positive thresholds (full lines) are those shown at upper positions, and the larger negative thresholds (dashed lines) are shown at lower positions. The dotted horizontal line has the sole purpose of separating the curves relative to positive (above the line) and negative (below the line) thresholds.}
\label{fig:osman}
\end{figure}
%%%%%%%%%%%%%%%%%%%%%%%%%%%%%%%%%%%%%%%%%%%%%%%
The 12 panels of Figure~\ref{fig:osman} show the conditioned average profiles obtained using the HVM simulation results, for some of the physical quantities described in Section~\ref{sec:numres} and indicated in each panel. The average has been computed up to a distance $\delta=12d_p$ from the LET peaks.
For each panel, the different line styles refer to the different threshold values, as listed at the bottom of the figure. All curves have been vertically shifted for clarity, and are therefore given in arbitrary units. 
The dotted horizontal line has the only purpose of separating the positive from the negative threshold values.
The figure shows that most of the variables have a clear enhancement at $\delta=0$ in the positive LET peak cases, which is more evident for larger threshold values. At the same time, in most cases the negative LET peaks are associated with a reduction of the conditioned average, again more evident for larger negative thresholds. This observation generally suggests that kinetic processes are more effective within a few $d_p$ from the intermittent structures, or more precisely from the locations where the fluid-scale turbulent cascade has produced a larger flux of energy.

A more detailed look at the individual panels of Figure~\ref{fig:osman} can provide information that may help uncovering the kinetic processes occurring in the simulation.  Panel A shows that the off-plane current $j_z$ is not well co-located with the positive LET structures, as the conditioned average profiles show a weak peak near $\delta\simeq 3d_p$. On the contrary, regions of strong negative LET are associated to smaller current, as evident from the bottom part of panel A. This could arise from the fact that negative LET structures are associated with regions where the turbulent production of small scales is locally inhibited, indicating possible equilibrium-like patterns, that results from a process of suppression of nonlinearity \citep{ServidioEA08}.

In panels B and C, the average proton and $\alpha$ particle vorticity peak at large positive LET locations, and, as for the current, is minimal for large negative LET. The important correlation between vorticity structures and kinetic features has been already observed in particle-in-cell simulations~\citep{parashar16}. 

Panel D shows that the exchange of energy between particles and fields, represented here through the quantity $\bE\cdot\bj$, is weakly enhanced near the LET positive peaks, but is relatively unaffected by the presence of negative LET peaks. This is in good agreement with previous works, where it has been shown  that locally the cascade produces narrow heating and dissipation patterns~\citep{KarimabadiEA13}.

Panels E and F display enhanced temperature near the positive LET peaks for both species, which extends up $\delta \sim $ 2--3$d_p$ from the peaks. For the $\alpha$ particles, there is indication of colder temperature near the large negative LET points, which highlights the possible presence of a differential heating processes of the two species~\citep{kasper08,maruca12}.

Panels G and H show the conditioned profiles of the temperature anisotropy $A_s$.
In panel G, the proton temperature anisotropy is only weakly affected by large positive energy transfer, while a general anisotropy increase can be observed in correspondence with large negative LET. 
On the contrary, as can be seen in panel H, the $\alpha$ particle temperature anisotropy increases (decrease) near the large positive (negative) LET structures. In this case, the LET dependence is similar for the temperature anisotropy and for the total temperature. This suggests that the presence of large turbulent energy transfer increases predominantly the perpendicular temperature, indicating for example the possible role of ion-cyclotron or mirror instabilities~\citep{matteini13}.

Panels I and J show that the quantity ${\mathbf{b}} \cdot \omega_s$ has similar conditioned averages as $\omega_{zs}$, for both protons and $\alpha$ particles. 

Finally, panels K and L show that the parameter $\xi_s$, quantifying the general deviation from Maxwellian VDF, is clearly larger at both positive and negative LET structures, and for both ion species. This last observation confirms that the kinetic processes resulting in modification of the ion VDFs are predominantly seen in the proximity of the locations with enhanced turbulent energy flux, where free energy is available for small-scale processes to occur. 
Additionally, it suggests that regions characterized by absent or reduced direct energy transfer towards small scales (large negative LET) are also characterized by the presence of robustly non-Maxwellian VDF. Whether this indicates that kinetic effect may provide feedback to the larger scales through some form of inverse cascade remains an open question that deserves further analysis.

It is interesting but not surprising to note that, for most of the physical quantities studied here, the observed peaks extend to approximately 2--3$d_p$ around the LET structures. This indicates that the effects of the kinetic features are typically smeared out within such range.

\section{Conclusions}
\label{sec:concls}

Exploring the details of the cross-scale connections between the MHD turbulent energy cascade and the small-scale kinetic processes is an important key to understand the role of dissipation mechanisms in collisionless plasmas~\citep{crossscale,alexandrova13,chen16}. Numerical simulations can provide important insight on general processes, in particular when experimental systems and observations are not easily controlled.

In this article, we have presented the analysis of one snapshot in the quasi-steady state turbulence of a two-dimensional Hybrid Vlasov-Maxwell numerical simulation (HVM), with the inclusion of $\alpha$ particles. The range of scales that have been investigated includes both the fluid range, where MHD turbulence transfers energy across scales, and the sub-ion range, where kinetic effects dominate the dynamics. This multi-scale investigation allows to explore the role of the turbulent energy that reaches the typical ion scales in the form of highly localized structures and the kinetic processes occurring at small scales. The former is represented by means of a proxy of the Local Energy Transfer rate (LET), whose heuristic definition is based on the linear scaling laws of the mixed third-order moment of the MHD fluctuations. The latter are described here through several quantities that can be related to kinetic processes, as for example the ion temperature and its anisotropy with respect to the magnetic field direction, or an index of deviation from Maxwellian particle velocity distribution. 

We have initially described the correlations between such kinetic quantities, first by visually comparing their maps, then by looking at some examples of joint probability distributions of such quantities pairs, and finally by estimating the nonlinear correlation coefficients between all the pairs. Several diagnostics show significant correlation. For example, the ion vorticity and the associated temperature show moderate correlation for both species, confirming previous observations that the temperature is enhanced in correspondence of vorticity structures~\citep{servidio15,franci16,valentini16,parashar16}. The deviation from Maxwellian~\citep{servidio12,valentini16} is also very well correlated with the pressure tensor anisotropy, revealing its relevant contribution in determining such deviation.

We have then analyzed the statistical properties of the energy transfer rate in HVM turbulence. The domain-averaged third-order moment suggests that the turbulent cascade might be fully developed in the range approximately between $\sim 2 d_p$ and $\sim 10 d_p$, where the measured energy flux is compatible with linear scaling. A direct cascade of energy is present (the negative sign of the third-order moment $Y$), in agreement with the expectations for turbulence in the MHD inertial range of scales. However, looking at the space-scale distribution of the local contributions to such scaling law, it is clear that both positive and negative energy fluxes exist, and are in comparable quantity, all across the simulation domain, the negative sign being the global balance between those. 

The scaling properties of the LET were modeled in terms of stretched exponential distributions, which provided a satisfactory description of the PDFs at all scales. The imbalance between positive and negative values of $\epsilon$ are well described by the asymmetry of their distributions. Moreover, the shape parameter of the fitted distributions shows power-law scaling in the inertial range, in agreement with previous observations of solar wind turbulence~\citep{sorriso2018} and confirming the reliability of the proxy for the description of the turbulence properties.

Finally, we have explored the presence of correlations between  the LET and the quantities described before. To this aim, we have calculated the conditionally averaged profile of each of those quantity, around the locations where the LET exceeds some given thresholds~\citep{osman12,sorriso2018}. This analysis may provide more detailed information that can help understand which kinetic processes are responsible for the removal (or transformation) of the turbulent energy from the system. These profiles shows that most of the physical quantities related to ``dissipative'' or kinetic processes are enhanced near regions of large positive LET, indicating a large flux of energy towards small scales. Conversely, these are often reduced near strong negative LET, likely associated to energy being transferred towards large scales, and thus indicating less availability of energy in the kinetic range.

Our results suggest, for example, that peaks of turbulent energy transfer are associated with enhanced temperature. More in particular, the perpendicular temperature increases near the high LET locations, indicating the possible role of ion-cyclotron or mirror instabilities~\citep{matteini13}.  The typical size of the region where the enhancement is observed is, for most quantities, of the order of 2--3$d_p$, while at larger distance their persistent fluctuations are averaged out. It is also clear from the analysis that the vorticity is a better tracer of energy availability than the current density~\citep{servidio15,franci16,parashar16,valentini16}, and that correlations are more evident for the $\alpha$ particles than for protons~\citep{valentini16}. 
Finally, deviations from Maxwellian VDF are evident for both positive and negative peaks of LET, an interesting feature that deserves further analysis.

The study of the local properties of the turbulent cascade can be a valuable tool for the exploration of the kinetic range processes in space plasmas. The validation of its use in numerical simulations provided in this work suggests its possible use with the recently available high-quality plasma measurements in the solar wind and in the magnetosphere, provided by the Multiscale Magnetospheric mission MMS~\citep{mms}. Extension to more complete descriptions of the energy transfer rate could be also used, allowing, for example, to examine the actual role played by compressive effects~\citep{carbone,AndresEA18}.

\section*{Acknowledgements}
This work has been partly supported by the Agenzia Spaziale Italiana under the Contract No. ASI-INAF 2015-039-R.O ``Missione M4 di ESA: Partecipazione Italiana alla fase di assessment della missione THOR'' and by the International Space Science Institute (ISSI) in the framework of International Team 504 entitled ``Current Sheets, Turbulence, Structures and Particle Acceleration in the Heliosphere''. L.S.-V., D.P., S.S. and I.Z. acknowledge support from the Faculty of the European Space Astronomy Centre (ESAC). Numerical simulations here discussed have been run on the Newton parallel machine at the University of Calabria (Italy).

\bibliographystyle{jpp}

\end{document}